\documentclass[draftcls,onecolumn]{IEEEtran}
\usepackage{amsmath,amsfonts}
\usepackage{algorithmic}
\usepackage{algorithm}
\usepackage{array}
\usepackage{textcomp}
\usepackage{stfloats}
\usepackage{url}
\usepackage{verbatim}
\usepackage{graphicx}
\usepackage{cite}
\usepackage{subcaption} 
\usepackage[dvipsnames]{xcolor}
\begin{document}
\bstctlcite{BSTcontrol}
\title{Thermal Stability and Depinning Currents of Domain Wall-Based Artificial Synapses}

\author{Guntas Kaur, Tanmoy Pramanik,
\thanks{Authors are with Department of Electronics and Communication Engineering, Indian Institute of Technology Roorkee, Roorkee, India, 247667}
\thanks{This work has been submitted to the IEEE for possible publication. Copyright may be transferred without notice, after which this version may no longer be accessible.}}

\markboth{Journal of \LaTeX\ Class Files,~Vol.~14, No.~8, August~2021}%
{Shell \MakeLowercase{\textit{et al.}}: A Sample Article Using IEEEtran.cls for IEEE Journals}

\maketitle
\begin{abstract}
Micromagnetic modeling is employed to optimize the design of artificial synapse devices based on the spin-orbit-torque (SOT) driven domain wall (DW) motion along a nanotrack with triangular notches. Key attributes, such as the thermal stability of the pinned DW and depinning currents, are obtained for varied nanotrack geometry and pinning strength. Depinning probability as a function of SOT current density and pulse width is studied using finite temperature micromagnetic simulations for varying pinning potential. Results show that wider notches provide better thermal stability - depinning current trade-off. On the other hand, narrow notches exhibit less variation in the depinning times at finite temperatures. It is observed that the DW position can be set precisely to any desired location by the SOT current pulse if the thermal stability is sufficiently high. It is also observed that the meta-plastic functionality can be obtained by adding notches of progressively higher depinning currents along the rectangular nanotrack.
\end{abstract}


\section{Introduction}

\IEEEPARstart {A}{rtificial} neural networks (ANN) mimic the functionality of the human brain by using an interconnection of neuron and synapse-like devices similar to the biological network of neurons \cite{markovic_physics_2020}. ANNs are trained by optimizing the weights stored in the synapses that work as non-volatile memories with two or more states \cite{burr_neuromorphic_2017,upadhyay_emerging_2019,chakraborty_pathways_2020,duan_memristor_based_2024}. Over the past decade, a number of studies have explored domain wall based magnetic tunnel junction (DW-MTJ) devices for hardware implementation of artificial synapses \cite{sharad_spin_based_2012,siddiqui_magnetic_2020,misba_voltage_controlled_2022,liu_domain_2021,leonard_shape_dependent_2022,yadav_demonstration_2023,alamdar_domain_2021,kumar_ultralow_2023}. In a typical DW-MTJ device, a current pulse of suitable amplitude and duration is used to set the location of the DW within the nanotrack, thanks to the spin-orbit-torque (SOT) driven DW motion \cite{emori_current_driven_2013,lee_position_2023}. The position of the DW is then mapped to a conductance value by the MTJ structure with the nanotrack acting as the free layer magnet \cite{leonard_shape_dependent_2022}. The conductance values represent programmed weights in such an artificial synapse device when placed in a crossbar array \cite{sharad_spin_based_2012}. DW-MTJ-based artificial synapses have many favorable properties, e.g., linear conductance variation, meta-plastic behavior, and capability of storing hidden weights \cite{yadav_demonstration_2023, leonard_shape_dependent_2022, kumar_ultralow_2023}. As with the magnetic random access memory (MRAM) technology \cite{golonzka_mram_2018,gallagher_22nm_2019,naik_stt_mram_2021}, DW-MTJ devices are also expected to have high endurance, tunable data retention, and compatibility with silicon integrated circuit technology for mass manufacturing. Compared to the conventional MTJ used as a synapse, DW-MTJ offers multiple states to store the synaptic weights with higher precision \cite{goodwill_implementation_2022,jung_crossbar_2022}. Although the concept of DW-MTJ-based synapse devices has been studied for over a decade \cite{sharad_spin_based_2012}, prototype devices have been demonstrated only recently \cite{siddiqui_magnetic_2020,leonard_shape_dependent_2022,kumar_ultralow_2023}. These demonstrations show that the domain wall position can be precisely controlled with different pinning structures that are lithographically defined along the DW track \cite{leonard_shape_dependent_2022,kumar_ultralow_2023}. However, the dependence of thermal stability (relevant for data retention) and depinning currents on the device geometry remains unclear. For example, retention times of a few hours are reported in the recent studies \cite{leonard_shape_dependent_2022,kumar_ultralow_2023}, which is insufficient for a long term memory. Variation of stored weights due to poor retention is known to impact the inference accuracy of ANNs \cite{DNN_Neurosim_2019}. Moreover, no guidelines are known for optimizing such devices for write speed, energy consumption, and area efficiency. The aim of this study is to explore these aspects of the DW-MTJ synapse. 
\par In this work, we consider a rectangular nanotrack geometry for the DW-MTJ with triangular notches on both sides \cite{leonard_shape_dependent_2022}. The notches can pin the DW at predefined locations on the nanotrack, making them thermally robust at smaller device dimensions while providing a controlled quantization of the conductance values. Thermal stability factors and depinning currents at $T = 0 $ K are calculated for varying notch dimensions. We also show that the pinning effect created by the triangular notches can be represented by a Gaussian potential well. The Gaussian pinning potential also allows us to describe DW motion using a simplified one-dimensional (1D) collective coordinate model. The accuracy of the 1D model is then compared with the full micromagnetic simulations. Finite temperature micromagnetic simulations are carried out to understand the stochastic nature of the DW depinning. Next, a series of notches are added to study the synaptic functionality of the DW-MTJ. We establish a methodology to optimize the write current and pulse widths to precisely control the DW position in these multi-state devices. Finally, we show that by choosing specific notch dimensions, meta-plastic functionality can be realized, as shown in recent experiments \cite{leonard_shape_dependent_2022}. Our findings show that careful optimization can enable fast, energy-efficient, and well-repeatable DW movement in nanotracks much smaller than recent prototypes.
\section{Simulation Methodology}
\begin{figure}[!t]
\centering
\includegraphics[width=5in]{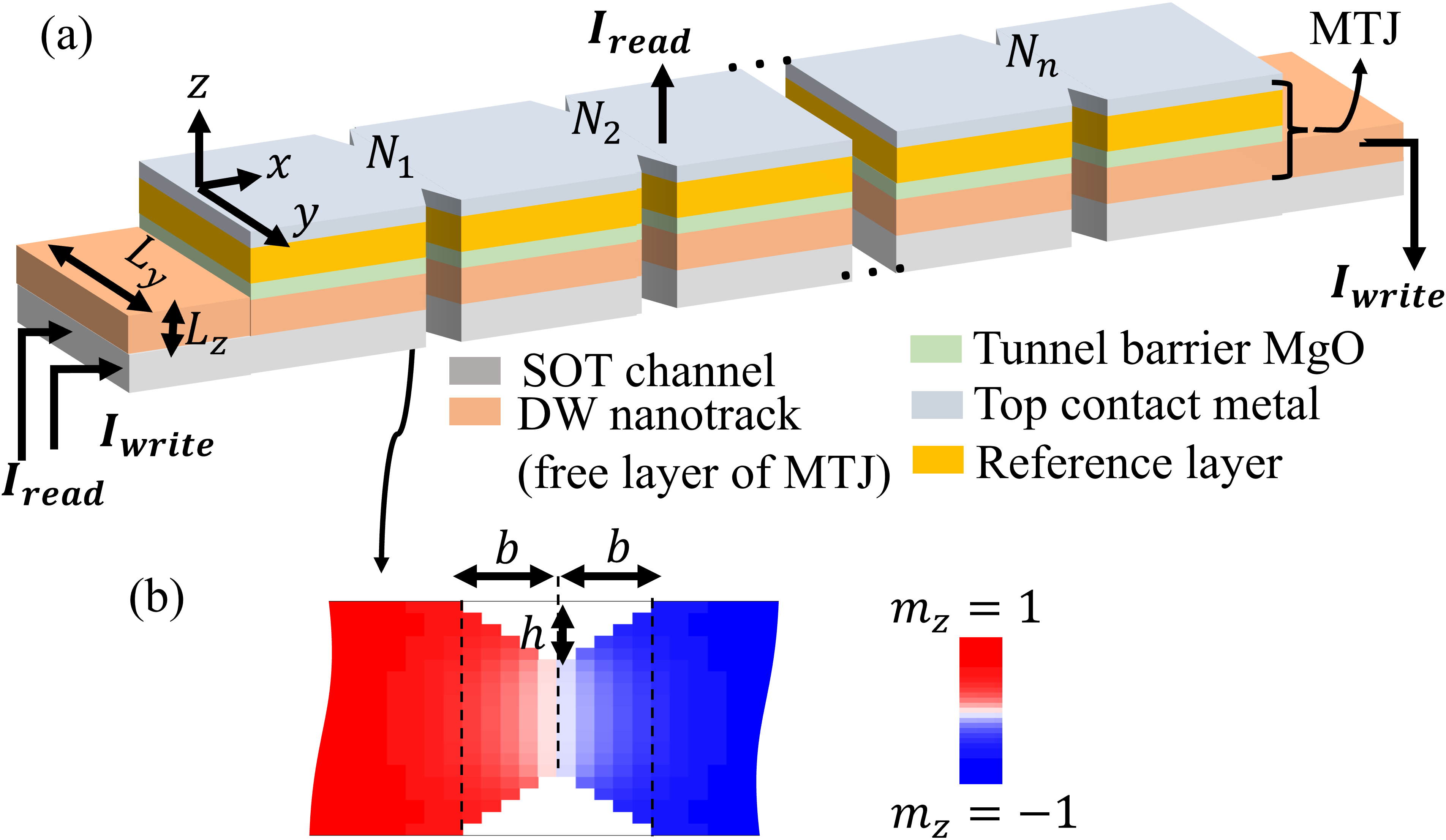}
\caption{Schematic of a typical DW-MTJ device: (a) shows the complete MTJ structure with all layers and $n$ notches marked as $N_1$, $N_2$, ..., $N_n$. The current paths for reading and programming are shown. Only the nanotrack layer is simulated in this study. (b) shows a DW pinned at one of the notches in the nanotrack. The color map corresponds to the perpendicular component of the magnetization unit vector, $m_z$. The parameters $b$ and $h$ define the triangular notch shape and control the pinning strength. Nanotrack width, $L_y$, and thickness, $L_z$, are marked in (a).}
\label{fig_1}
\end{figure}
The typical DW-MTJ-based synapse device is shown in Fig. \ref{fig_1}(a). We assume perpendicular magnetic anisotropy material in the DW nanotrack. The nanotrack containing the DW acts as the free layer ferromagnet in the MTJ stack. In this simplified setup, we have assumed the whole MTJ to have the same cross-section as the DW nanotrack. Note that, both ends of the nanotrack may need additional MTJ structures for practical implementation \cite{lee_position_2023,alamdar_domain_2021,raymenants_nanoscale_2021}. For this study, it suffices to simulate the DW track only, assuming a current flowing through the SOT channel will create spin-transfer torque on the FM nanotrack, causing the DW to move. The geometry and material parameters for the FM nanotrack are listed in Table I. We have chosen parameters typical to the CoFeB/MgO material system \cite{kumar_emulation_2022}. We also include the Dzyaloshinskii–Moriya interaction (DMI) energy term in our calculation that sets a preference for N$\mathrm{\acute{e}}$el type DW with left-handed chirality \cite{rohart_skyrmion_2013}. The magnetization configuration of the DW at rest is shown in Fig. \ref{fig_1}(b). The parameters $b$ and $h$ [see Fig. \ref{fig_1}(b)] define the notch shape, which controls how strongly the DW is pinned. All micromagnetic simulations are run using OOMMF \cite{oommf_2016}. To find the trapped DW's energy landscape and thermal stability factor, we have used the string method as detailed elsewhere \cite{chaves_oflynn_energy_2013}. For $T = 300$ K micromagnetic simulations, the relaxed DW solutions obtained at $T=0$ K are allowed to evolve for 15 ns without the SOT term to achieve thermal equilibrium solutions. SOT-driven motion is then simulated using the thermal equilibirum solution as the initial state. 
\par The 1D model describing DW depinning is implemented considering rigid DW motion along the nanotrack. Within the 1D model, we solve for the evolution of DW position, $X$, and the orientation of magnetic moments at the wall position, $\Phi$, using the equations \cite{martinez_current_driven_2014,moretti_micromagnetic_2017}: 
\begin{equation}
	\dot{X} = \frac{\delta}{1+\alpha^2} (\Omega_A + \alpha \Omega_B) 
\end{equation}
\begin{equation}
	\dot{\Phi} = \frac{1}{1+\alpha^2} (-\alpha \Omega_A + \Omega_B).
\end{equation}
In the equations above, we define $\Phi$ as the angle with the positive x-axis, i.e., $\Phi(t=0)=\pi$ for the left-handed N$\mathrm{\acute{e}}$el wall. $\alpha$ is the Gilbert damping factor. $\delta=\sqrt{A_{ex}/(K_u - 1/2(\mu_0 M_s^2))}$ is the DW width calculated from parameters as in Table I. $\mu_0$ is the vacuum permeability. The terms $\Omega_B$ and $\Omega_B$ in the above equations are given as:
\begin{equation}
	\begin{split}
		\Omega_A & = \frac{1}{2}\gamma[-H_k \sin(2 \Phi) + \pi H_D \sin \Phi]
	\end{split}
\end{equation}
\begin{equation}
	\Omega_B=\gamma \left[H_P + \frac{\pi}{2} H_{SH}\left[ \frac{L_y - 2h}{W(x)} \right] \cos \Phi\right]
\end{equation}
Where $\gamma$ is the gyromagnetic ratio, $H_k =\mu_0 M_s (N_{yy}-N_{xx})$ is the shape anisotropy field at the domain wall position, calculated assuming a rectangular box of sides $(\delta,L_y-2h,L_z)$, $H_D = D / (\mu_0 M_s \delta)$ is the DMI effective field, $H_{SH} = (\hbar\theta_{SH} J_c)/(2 \mu_0 qM_sL_z)$ is the SOT effective field including the charge current density $J_c$. $H_p(x) = [-1/ (2 \mu_0 M_sL_yL_z)] \partial V_{\mathrm{pin}}(x)/ \partial x$ is the effective pinning field created by the notch shape, where the potential $V_{\mathrm{pin}}(x)$ is the potential energy describing the pinning effect. $W(x)$ takes into account the varying width of track due to notches placed on both sides. 
\par In the following sections, we first discuss the results for a nanotrack with a single notch placed in it. Next, we consider a nanotrack with multiple notches of the same pinning strength (i.e., same $h$ and $b$) placed at a distance such that their pinning potentials do not interfere. Finally, we present results where multiple notches of different pinning strengths are considered.

\begin{table}[!t]
\caption{Simulation parameters\label{tab:table1}}
\centering
\begin{tabular}{|c|c|}
\hline
Parameter & Value\\
\hline
Width of the DW track ($L_{y}$) & 40  nm\\
\hline
Thickness of the DW track ($L_{z}$) & 1  nm\\
\hline
Saturation magnetization ($M_{s}$) & 1000 kA/m\\
\hline
Exchange coefficient ($A_{ex}$) & 15 pJ/m\\
\hline
Saturation Magnetization ($\alpha$) & 0.012\\
\hline
Spin-Hall angle ($\theta_{\text{SH}}$) & 0.4\\
\hline
Perpendicular anisotropy energy ($K_{u}$) & \(10^6 \, \text{J/m}^3\) \\
\hline
DMI coefficient ($D$) & \(0.5\,\text{mJ/m}^2\)\\
\hline
Temperature ($T$) & 0 K, 300 K\\
\hline
\end{tabular}
\end{table}

\section{Results}
\subsection{Estimation of thermal stability factor and pinning potential}
\begin{figure}[bt]
\centering
\includegraphics[width=5in]{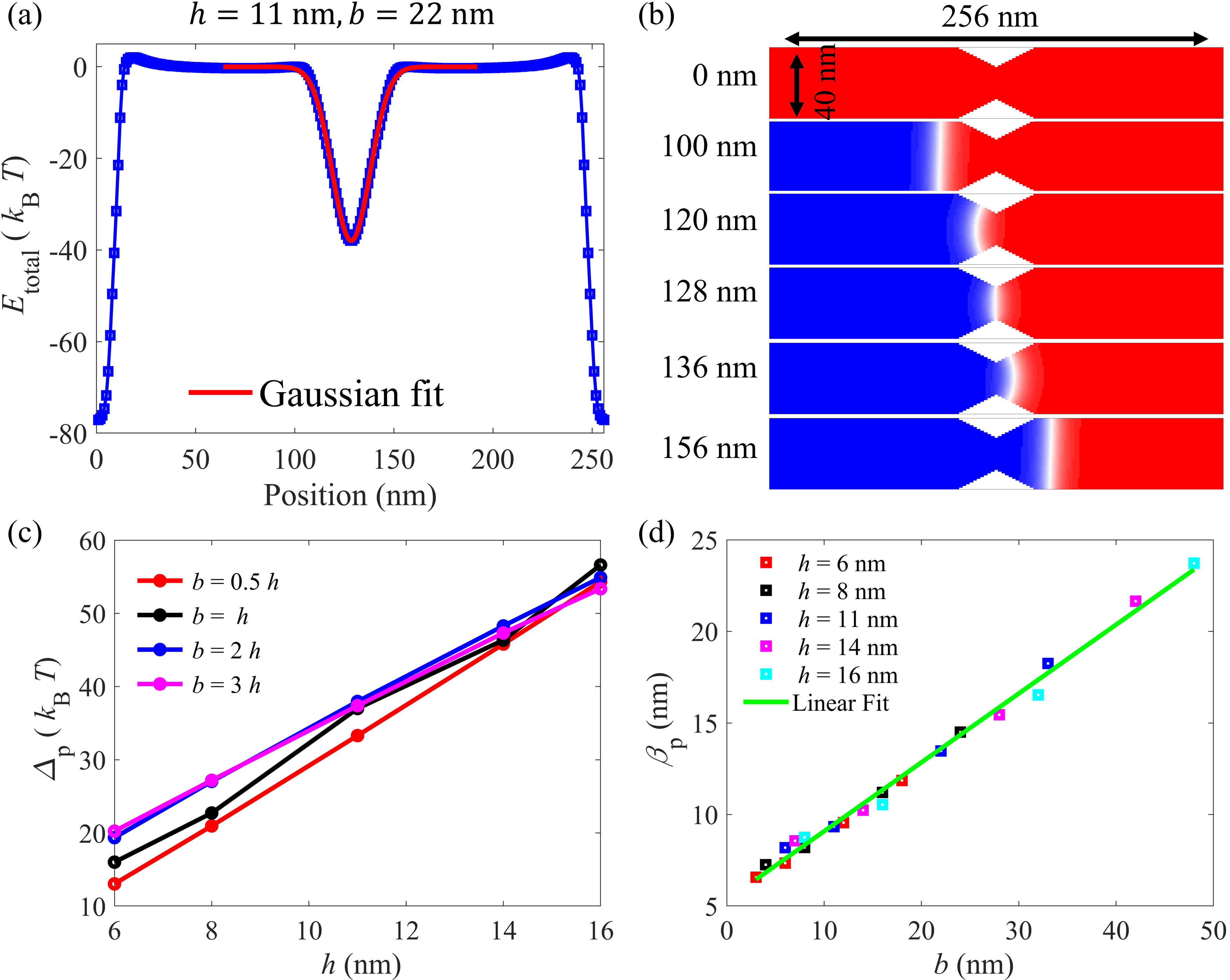}
\caption{(a) Total energy profile calculated from the string method, as the DW moves through a notch. The global energy minima at both ends (0 and 256 nm) denote uniformly magnetized states. The notch is placed at the center position of 128 nm. Dimensions are noted in the figure. The local energy minimum created by the notch can be fit to a Gaussian potential (red line). Snapshots of out-of-plane magnetization ($m_z$) for selected positions in (a) are shown in (b). The color map is the same as in Fig. \ref{fig_1}. Fit parameters $\Delta_p$ and $\beta_p$ for varying ($h$, $b$) are shown in (c) and (d), respectively. $\Delta_p$ indicates the depth of the potential well and directly gives the thermal stability factor of the pinned DW state. $\beta_p$ indicates the shape of the potential well and is directly proportional to $b$.}
 \label{fig_2}
\end{figure}

Compared to a DW track with a uniform width, the notched structure provides a lower energy position for the DW, keeping it trapped unless a magnetic field or current is applied that can overcome the pinning strength. The potential well created by the notch could be estimated by the difference between the total DW energy at the narrowest part of the notch and the widest part of the track \cite{EDTM_Guntas}. Here, we have calculated this potential profile employing the string method. Figure  \ref{fig_2}(a) shows the typical potential energy profile obtained for a single notch in a nanotrack of width $L_y=40$ nm. The snapshots in Fig. \ref{fig_2}(b) show the magnetization configuration in the nanotrack for selected positions of the DW. The total magnetic energy has a local minimum when the DW is at the center of the notch (located at a position of 128 nm). For the small dimensions of the nanotrack (40 nm $\times$ 256 nm) considered here, the global minimum energy configurations are the uniformly magnetized states (endpoints of the string at 0 nm and 256 nm). Hence, the potential well at the notch must be deep enough to retain the DW for the device's lifetime. Following the usual guidelines for MRAM, it is expected that a potential well depth of $ 60k_BT $ or higher would be sufficient to hold the DW at the notch position.    
\begin{figure}[tb]
\centering
\includegraphics[width=5in]{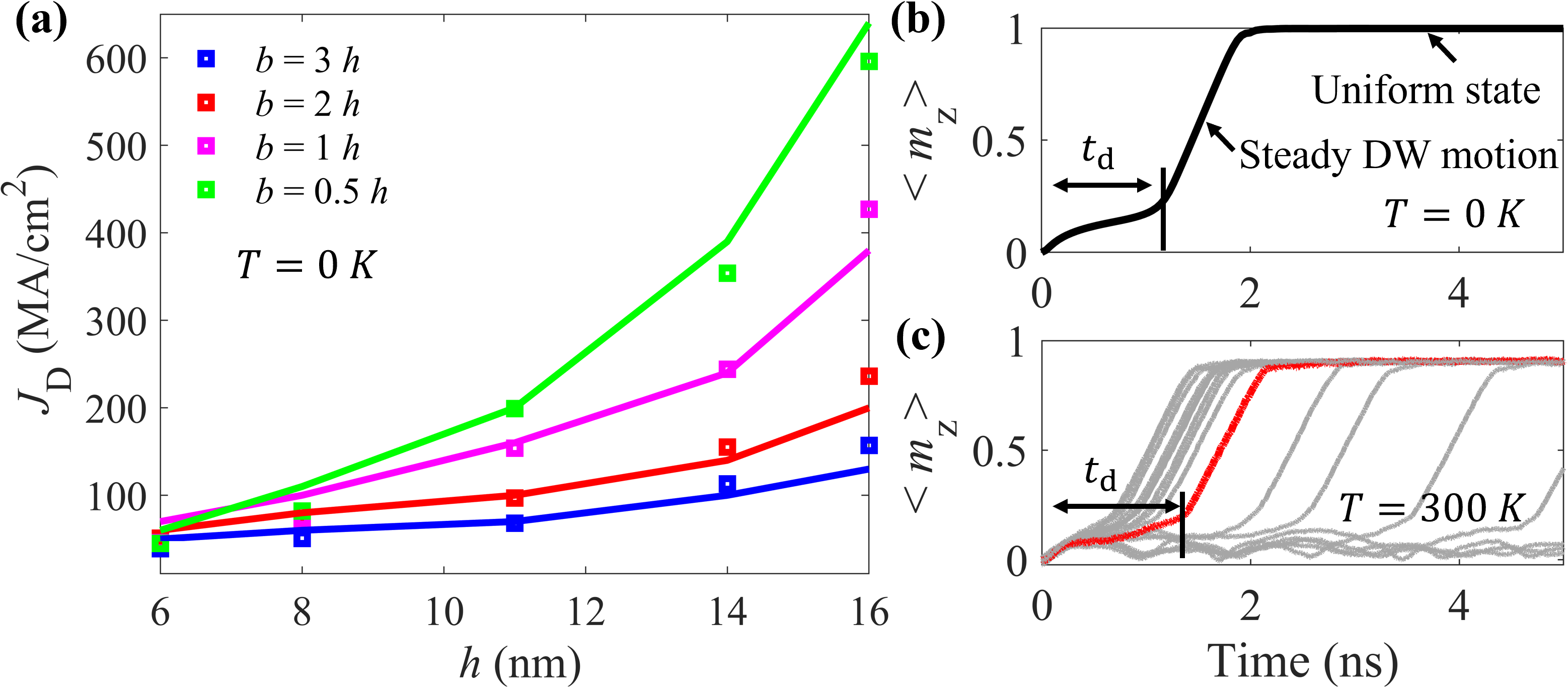}
\caption{(a) Variation of depinning current density, $J_D$ with  $h$ and $b$ at $T=0$ K. For a given $h$, higher $b$ lowers $J_D$. Solid lines are obtained from the 1D model. Squares are obtained from the micromagnetic simulation. (b) shows the trajectory of space averaged out-of-plane magnetization, $\langle m_z\rangle$, as the DW is depinned. (c) shows the stochastic nature of DW motion and depinning time at $T=300$ K. Each grey line is one independent micromagnetic simulation. Depinning time, $t_\mathrm{d}$, marks the time needed for the DW to come out of the notch.}
  \label{fig_3}
\end{figure}
\par To extract the pinning potential, $V(x)$ from the total energy, we fit the total energy profile near the local minima with a Gaussian form given as:
\begin{equation}
	V(x) = \Delta_p exp[-(x-\alpha_p)^2/\beta_p^2]\
 \label{single_gaussian}
\end{equation}
Where $\Delta_p$, $\alpha_p$ and $\beta_p$ are fitting parameters. The parameter $\alpha_p$ is simply the position of the notch, reducing the number of fitting parameters to two ($\Delta_p$ and $\beta_p$). The red line in Fig. \ref{fig_2}(a) shows the Gaussian fit around the local energy minima. Note that the parameter $\Delta_p$ in the above equation denotes the depth of the potential well, which determines the thermal stability of the pinned DW. The parameter $\beta_p$ indicates the width of the well and is proportional to the notch width parameter $b$. The variation of $\Delta_p$ and $\beta_p$ is shown in Fig. \ref{fig_2}(c) and (d), respectively, for varying notch dimensions ($h$, $b$). As the notch depth $h$ increases, both the pinning strength and thermal stability increase, which is evident from Fig. \ref{fig_2}(c). Once the fitting parameters $\Delta_p$ and $\beta_p$ are known, they can be used in the 1D model to specify the potential $V_{pin}(x)$ as described above.
\subsection{Depinning currents at $T = 0$ K}
\par Next, we calculate the the minimum current required to depin the DW from the energy minima at the center of the notch. Note that the current density would be higher at the notch position than at the wider part of the nanotrack for a given current flowing through the SOT channel. The current density mentioned here corresponds to the value at the center of the notch. For the $T=0$ K simulations, we have applied progressively higher current density, starting from a value of $10 MA/cm^2$ with a step of $1 MA/cm^2$. The minimum current density that destabilizes the DW is marked as the depinning current, $J_D$. The variation of $J_D$ at $T = 0$ K for different $h$ and $b$ values is shown in Fig. \ref{fig_3}(a). The solid lines are obtained from the 1D model, and the squares are obtained from the micromagnetic simulations. The agreement of both models is evident. Figure \ref{fig_3}(a) shows increasing $J_D$ for stronger pinning strength (increasing $h$). Interestingly, we find that wider notches (higher $b$) allow us to lower the current density significantly without degrading the thermal stability [see Fig. \ref{fig_2}(c)]. Figure \ref{fig_3}(b) shows a trajectory of average out-of-plane magnetization, $\langle m_z \rangle$, during depinning followed by steady DW motion till the DW is driven out of the track at $\langle m_z \rangle\approx 1$. The time to drive the DW out of the notch is shown as the depinning time, $t_\mathrm{d}$. 
\subsection{Stochastic nature of depinning at $T = 300$ K}
\begin{figure}[tb]
\centering
\includegraphics[width=5in]{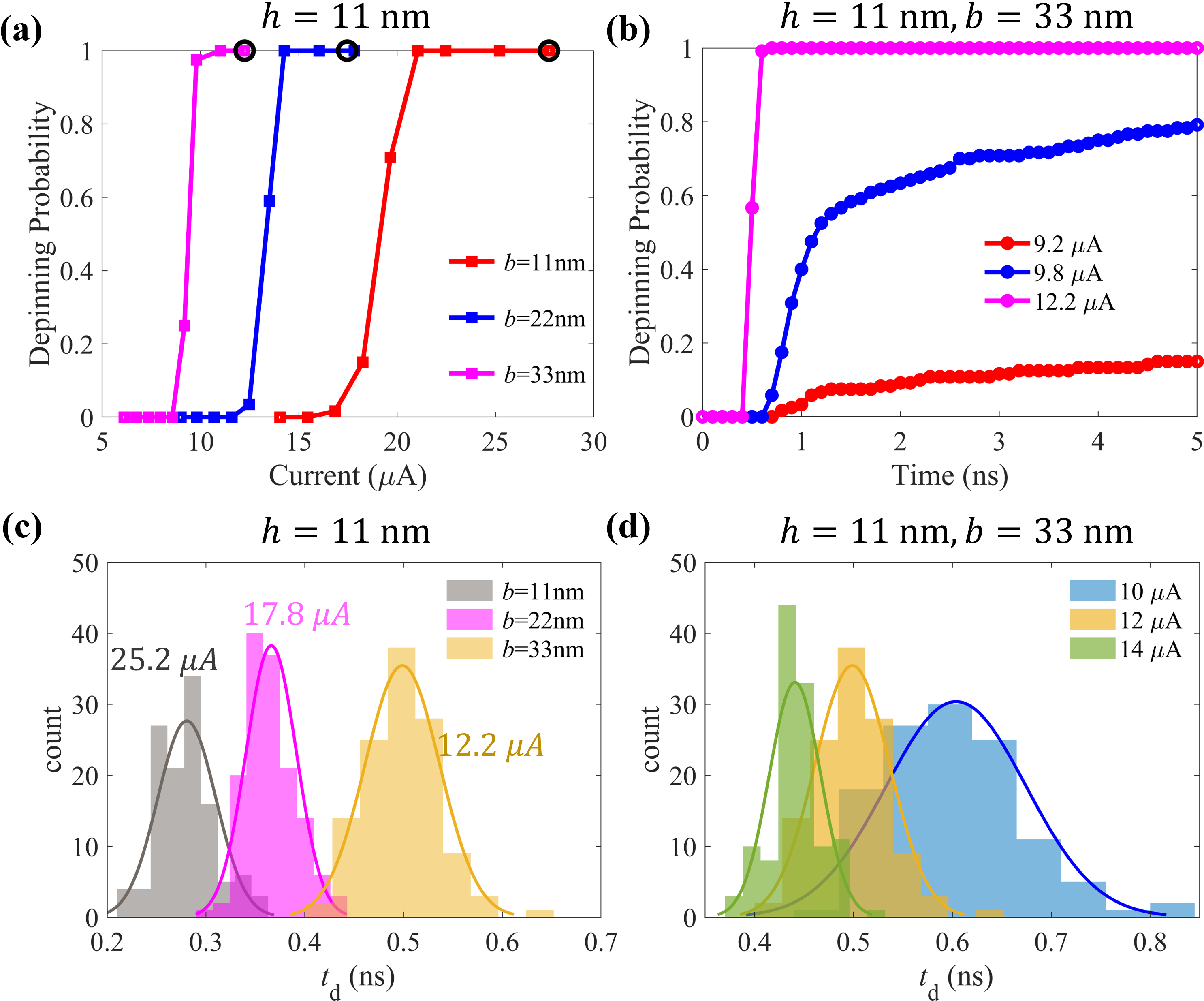}
\caption{The stochastic nature of DW depinning at $T=300$ K: (a) shows the dependence of depinning probability with the applied current for varying $b$, calculated from 144 simulations. The black circles mark the corresponding depinning currents at $T=0$ K. (b) shows the depinning probability increases steeply and then saturates, even if the current is left ON. (c) shows the distribution of depinning times ($t_\mathrm{d}$) for varying $b$. Minimum current achieving 100\% depinning is applied in each case. Both the average, $\mu(t_\mathrm{d})$, and the standard deviation, $\sigma(t_\mathrm{d})$, increase with $b$. For a given notch dimension, a higher current can suppress the thermal noise, lowering both $\mu(t_\mathrm{d})$ and $\sigma(t_\mathrm{d})$, as observed in (d).}
 \label{fig_4}
\end{figure}

The domain wall motion becomes stochastic once the random thermal field is turned on at 300 K. Figure \ref{fig_3}(c) shows multiple trajectories calculated from 300 K micromagnetic simulations. At $T=300$ K, DW can be depinned at a current density lower than the critical value $J_D$ described above.  As observed from Fig. \ref{fig_3}(c), there is a finite depinning probability, depending on the current amplitude and duration. Also, the depinning time, $t_\mathrm{d}$, is now different for each case [$t_\mathrm{d}$ is marked for the red trajectory in Fig. \ref{fig_3}(c)] and shows considerable variation. Interestingly, once the DW is depinned, DW velocity remains the same for all cases. Hence, there is hardly any variation in the propagation time of DW through the track. To understand the variation of the depinning probability and depinning times on the parameters ($h$, $b$), we have varied the current density around $J_D$ [Fig. \ref{fig_3}(a)], and 144 simulations are carried out for each case. Figure \ref{fig_4}(a) shows the depinning probability as a function of current for three different notch parameters. The black circles mark the corresponding depinning currents, $J_D$, at $T=0$ K. Figure \ref{fig_4}(b) plots the rate of depinning for varying applied currents for the case of $h=$11 nm, $b=$ 33 nm. The depinning occurs very quickly after the current is turned ON, and the depinning probability improves only a little even if the current is kept ON for extended periods. 
\par The variation of $t_\mathrm{d}$ is explained in Fig. \ref{fig_4}(c) \& (d). We observe that the wider the notch (increasing $b$), the higher the average depinning time, $\mu(t_\mathrm{d})$, and the standard deviation, $\sigma(t_\mathrm{d})$. Again, for the same notch dimension  ($h$, $b$), increasing the current amplitude lowers both $\mu(t_\mathrm{d})$ \& $\sigma(t_\mathrm{d})$ as seen in Fig. \ref{fig_4}(d).
\begin{figure}[tb]
\centering
\includegraphics[width=5in]{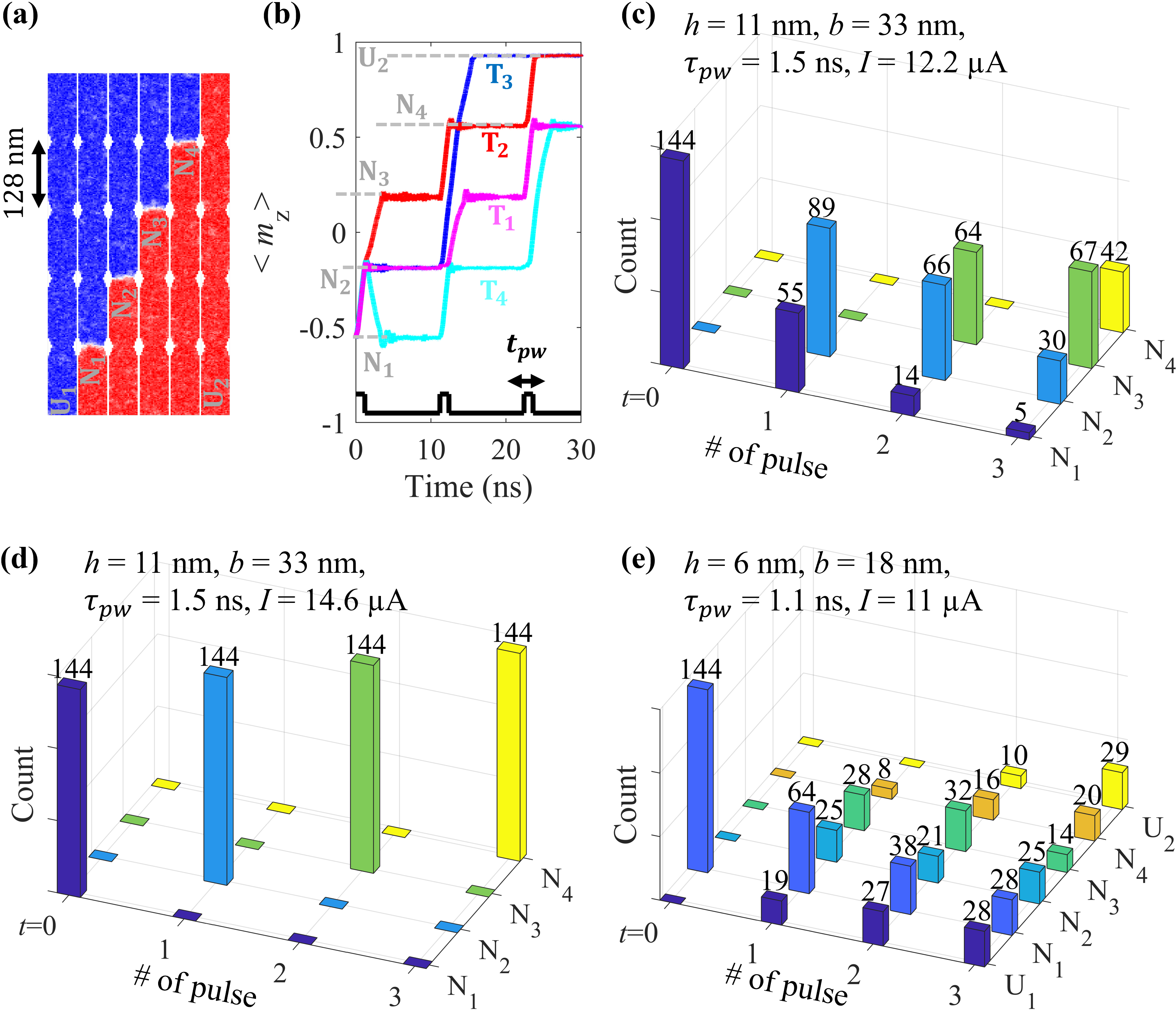}
\caption{Programming DW-MTJ synapse with four notches: (a) shows maps of $m_z$ and DW location in the track for different states. Uniformly magnetized states, $U_1$ and $U_2$ are also shown. (b) shows typical trajectories of $\langle m_z \rangle$ vs. time when current pulses are applied to the synapse, initialized with a DW at the first notch ($N_1$) at $t=0$. $T_1$ shows the desired response, where each pulse depins the DW and moves it to the next notch. $T_2$ shows a trajectory where the DW skips $N_2$ during the first pulse and settles directly at $N_3$. $T_3$ shows DW skipping multiple states during the second pulse. $T_4$ fails to depin at the first pulse. (c)-(e) show how precisely DW position can be controlled with three current pulses. 144 simulations are carried out for each condition. (c) shows many cases where the DW fails to depin, causing incorrect state after the pulse. (d) shows the desired response with an optimized current pulse. The optimization does not work if $\Delta_p$ is too low, as shown in (e). In the case of (e), the DW can easily escape from the track, and many instances of  $U_1$ or $U_2$ states are also visible.}
 \label{fig_5}
\end{figure}
\begin{figure}[tb]
\centering
\includegraphics[width=5in]{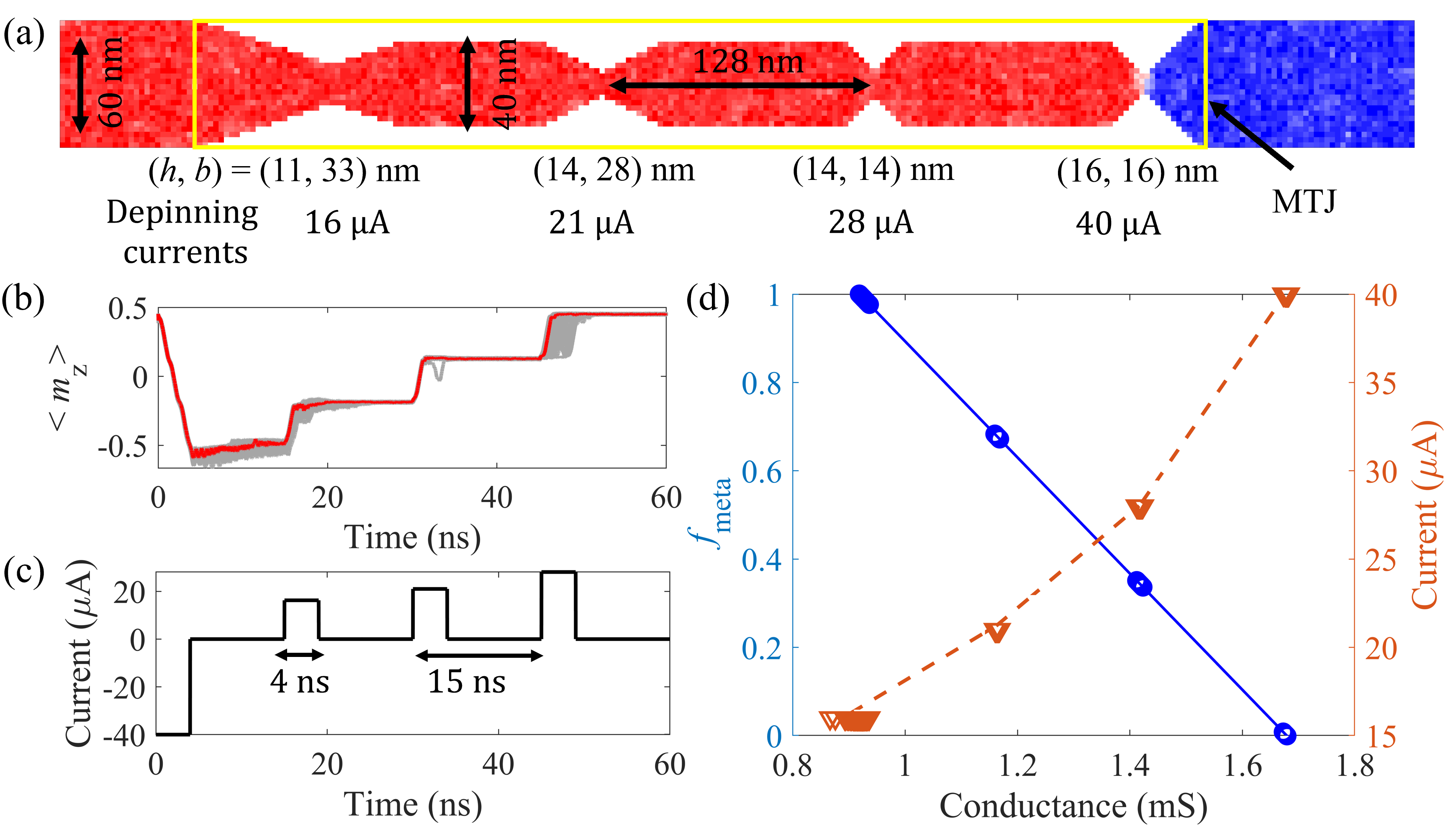}
\caption{DW-MTJ-based synapse with meta-plastic functionality: (a) shows the DW track with 4 notches of increasing pinning strength from left to right. MTJ coverage area is also marked. For this case, we have considered higher width at both sides to stop the DW moving out the track. Color map is same as Fig. \ref{fig_1}. (b) shows DW moving one step at a time when moving to the right and one step switching when moving to the left. Each gray line corresponds to one simulation out of 144. The red line highlights a typical trajectory. Corresponding current pulses are shown in (c). (d) shows the depinning current amplitude vs. conductance of the synapse on the right axis. Dashed lines are guide to the eye. The left axis shows a linear mapping of the conductance to a meta-plastic function, $f_{meta}$. Tunnel magneto-resistance ratio, TMR $=$ 150\% and resistance-area product for parallel orientation, $\mathrm{RA_P}=$ 10 $\mu m^2$ are used to calculate the equivalent conductance.}
 \label{fig_6}
\end{figure}
\subsection{Nanotrack with multiple notches: DW-MTJ-based synapse}
Next, a DW-MTJ-based synapse with four notches is studied to understand how the thermal stability factor and the probabilistic depinning impact synapse programming. The DW track with four notches of the same dimension ($h$, $b$) is shown in Fig. \ref{fig_5}(a). Current pulses are applied through the SOT channel during the programming step, setting the DW to the desired notch positions ($N_1$ through $N_4$). To study the write (or program) operation, we initialize the DW at the first notch ($N_1$), then apply a current pulse as shown in Fig. \ref{fig_5}(b). Again, to observe the stochastic variation, simulations are repeated 144 times. For each set of ($h$, $b$), we choose the current pulse amplitude corresponding to the depinning probability of 100\% from Fig. \ref{fig_4}. ON time, $t_\mathrm{pw}$ is varied while the OFF time is set to 10 ns to allow the DW to settle down after each pulse. The position of the DW is recorded at the end of each pulse. We find that if the current pulse amplitude and ON times ($t_\mathrm{pw}$) are not optimized, DW may move randomly and may even escape from the track, resulting in uniformly magnetized states ($U_1$ or $U_2$). Such cases are explained in Fig. \ref{fig_5}(b). The trajectory $T_1$ in Fig. \ref{fig_5}(b) is the desired case, where the DW moves precisely one step to the next notch with each pulse. For all the other cases, the DW either moves more than one step during a single pulse ($T_2$, $T_3$) or fails to move ($T_4$). We find that the optimum $t_\mathrm{pw}$ can be set considering the fastest [$t_{\mathrm{pw, fastest}}=\mu(t_\mathrm{d})-\sigma(t_\mathrm{d})$] and slowest [$t_{\mathrm{pw, slowest}}=\mu(t_\mathrm{d})+\sigma(t_\mathrm{d})$] depinning events, since the propagation times do not vary considerably [see Fig. \ref{fig_3}(c)]. The difference, $t_{\mathrm{pw, slowest}} - t_{\mathrm{pw, fastest}}=2\sigma(t_\mathrm{d})$, must be less than $\mu(t_\mathrm{d})$, to stop the fastest depinning event skipping a state. This condition indicates that lower $\sigma(t_\mathrm{d})$ is required to find an optimum $t_\mathrm{pw}$. The effectiveness of this optimization is revealed in Fig. \ref{fig_5}(c) \& (d) for the case of ($h=11$ nm, $b=$ 33 nm). With current pulse amplitude $I= 12.2$ $\mu A$, $\sigma(t_\mathrm{d})$ remains high, and an optimum $t_\mathrm{pw}$ does not exist, resulting in missed switching [Fig. \ref{fig_5}(c)]. These failed switching events are similar to $T_4$ of Fig. \ref{fig_5}(b). With increased current, $I= 14.6$ $\mu A$, clean step-by-step programming of DW is obtained [Fig. \ref{fig_5}(d)]. However, if the pinning strength of the notch is not sufficient (typically $\Delta_p<30 k_BT$), no optimum $t_\mathrm{pw}$ exists, and the programming becomes completely random, as shown in Fig. \ref{fig_5}(e).
\par Finally, we show that by choosing varying notch sizes, the rectangular DW-MTJ device can mimic the response of a trapezoidal DW-MTJ as reported in \cite{leonard_shape_dependent_2022}. A trapezoidal DW-MTJ with hidden weights and meta-plastic weight adjustment properties has been shown to perform better than the usual rectangular DW-MTJ with the same pinning strengths at all the notch positions \cite{leonard_shape_dependent_2022}. An example configuration of this modified nanotrack is shown in Fig. \ref{fig_6}(a). We have placed the notches with increasing depinning currents following Fig. \ref{fig_3}(a). Additionally, both ends of the track is made wider to prevent the DW from spilling over. Figure 6(b) shows the change in the DW position in forward and reverse directions when the current pulses of varying amplitude are applied. The sequence of current pulses with variable amplitudes are shown in Fig. \ref{fig_6}(c). The DW is initialized on the right most position as shown in Fig. \ref{fig_6}(a). On application of the first pulse (negative value), the DW moves to the left (decreasing $\langle m_z\rangle$). For the subsequent pulses (positive), the DW moves to the right(increasing $\langle m_z\rangle$). The switching is simulated 144 times to verify the repeatability. The behavior in Fig. \ref{fig_6}(b) is similar to the one reported for the trapezoidal structure \cite{leonard_shape_dependent_2022}. Figure \ref{fig_6}(d) shows a mapping of the DW position to conductance and a meta-plastic function, $f_{\mathrm{meta}}$. Compared to the area of the trapezoidal synapse (approximately $ 0.5\mu m\times 2\mu m$\cite{leonard_shape_dependent_2022}, the proposed synapse in Fig. \ref{fig_6} takes up only $ 0.06\mu m\times0.64\mu m$ area. 

\section{Conclusion}
In summary, our study has revealed the crucial role of device dimensions on DW's pinning strength and thermal stability in a rectangular nanotrack, which is essential for the repeatable and precise programming of DW-MTJ-based synapses. The results demonstrate that careful tuning of the notch dimension could enable sufficient thermal stability and significantly reduce depinning currents. At the same time, precise control of the DW position is possible if the current pulse amplitude and duration are correctly optimized. We have also established a methodology to describe the pinning potential in a 1D model of DW motion, which could be useful for developing compact models. By adding progressively varying pinning strength along a rectangular nanotrack, we are able to show meta-plastic functionality at a much smaller device dimensions.  These findings could provide a foundation for designing efficient and scaled DW-synapse devices with applications in neuromorphic computing.

\section*{Acknowledgment}
The authors would like to acknowledge the Supercomputing facility of IIT Roorkee, established under the National Supercomputing Mission (NSM), Government of India and supported by the Centre for Development of Advanced Computing (CDAC), Pune, for providing high-performance computing resources. T.P. acknowledges support from the Science and Engineering Research Board, Govt. of India, Grant \# SRG/2021/000377, and Faculty Initiation Grant from IIT Roorkee. G. K. acknowledges support from the MHRD, Govt. of India.

\bibliographystyle{IEEEtran}

\bibliography{references}

\end{document}